\documentstyle[epsf]{mn}
\begin{document}

\title[The Milky Way's dark halo]
      {Two measures of the shape of the Milky Way's dark halo}

\author[R.P. Olling, M.R. Merrifield]
       {Rob P. Olling$^{1}$\thanks{E-mail: olling@astro.rutgers.edu},      
        Michael R. Merrifield$^{2}$\thanks{E-mail: Michael.Merrifield@Nottingham.ac.uk}  \\
$^1$Department of Physics and Astronomy, Rutgers University, PO Box
       849, Piscataway, NJ 08855, USA \\
$^2$ School of Physics and Astronomy, University of Nottingham,
       University Park, Nottingham, NG7 2RD}


\maketitle

\begin{abstract}

In order to test the reliability of determinations of the shapes of
galaxies' dark matter halos, we have made such measurements for the
Milky Way by two independent methods.  First, we have combined the
measurements of the over-all mass distribution of the Milky Way
derived from its rotation curve and the measurements of the amount of
dark matter in the solar neighborhood obtained from stellar kinematics
to determine the flattening of the dark halo.  Second, we have used
the established technique based on the variation in thickness of the
Milky Way's H{\sc i} layer with radius: by assuming that the H{\sc i}
gas is in hydrostatic equilibrium in the gravitational potential of a
galaxy, one can use the observe flaring of the gas layer to determine
the shape of the dark halo.

These techniques are found to produce a consistent estimate for the
flattening of the dark matter halo, with a shortest-to-longest axis
ratio of $q\sim 0.8$, but only if one adopts somewhat non-standard
values for the distance to the Galactic centre, $R_0$, and the local
Galactic rotation speed, $\Theta_0$.  For consistency, one requires
values of $R_0 \la 7.6\,{\rm kpc}$ and $\Theta_0 \la 190\,{\rm km}\,{\rm
s}^{-1}$.  The results depend on the Galactic constants because the
adopted values affect both distance measurements within the Milky Way
and the shape of the rotation curve, which, in turn, alter the inferred
halo shape.  Although differing significantly from the current
IAU-sanctioned values, these upper limits are consistent with all
existing observational constraints. 

If future measurements confirm these lower values for the Galactic
constants, then the validity of the gas layer flaring method will be
confirmed.  Further, dark matter candidates such as cold molecular gas
and massive decaying neutrinos, which predict very flat dark halos with
$q \la 0.2$, will be ruled out.  Conversely, if the Galactic constants
were found to be close to the more conventional values, then there would
have to be some systematic error in the methods for measuring dark halo
shapes, so the existing modeling techniques would have to be viewed
with some scepticism.

\end{abstract}


  \section{Introduction}
\label{sec:Introduction}

The speed at which spiral galaxies rotate remains relatively constant
out to large radii \cite{RFT80,aB81}, which implies that they contain
large amounts of dark matter.  The nature of this material remains
obscure, but one key diagnostic is provided by the shape of the dark
matter halo, as quantified by its shortest-to-longest axis ratio, $q =
c/a$.  The roundest halos with $q \sim 0.8$ are predicted by galaxy
formation models in which hot dark matter is dominant \cite{jP93}.
Cosmological cold dark matter simulations typically result in triaxial
dark halos \cite{WQSZ92} while the inclusion of gas dynamics in the
simulations results in somewhat flattened, oblate halos
\cite{nKjeG91,sUlM94,jD94} with $q = 0.5 \pm 0.15$ \cite{jD94}.
Models for other dark matter candidates such as cold molecular gas
\cite{PCM94} and massive decaying neutrinos \cite{dS90} require halos
as flat as $q \sim 0.2$.  Clearly, the determination of $q$ for real
galaxies offers a valuable test for discriminating between these
cosmological models.

\begin{figure}
\epsfxsize 1.00\hsize
 \epsffile{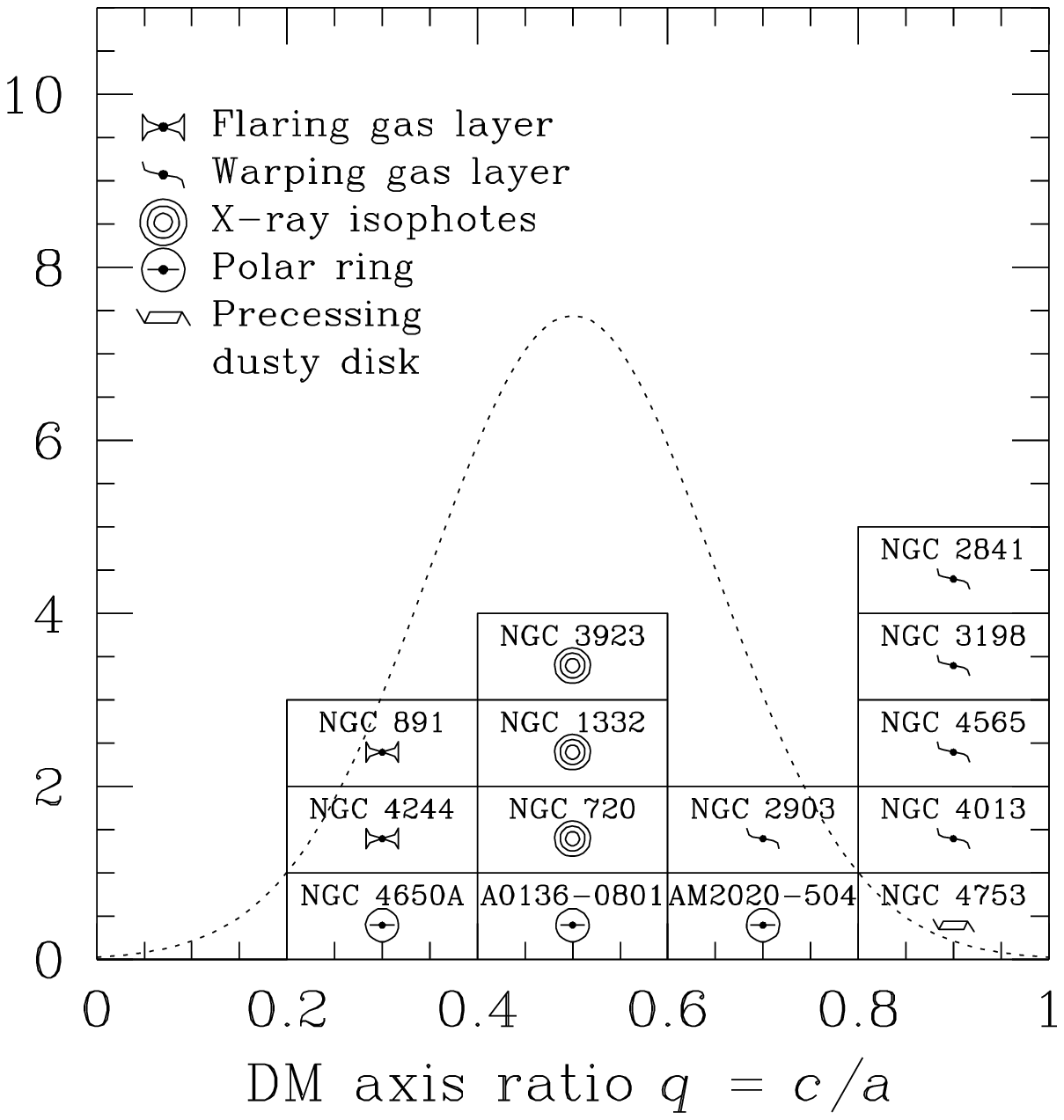}
\caption{ \label{fig:Halo_Shapes_cmp}A summary of the existing
estimates of galaxies' dark matter halo shapes as parameterized by
their shortest-to-longest axis ratio, $q$.  For each
individual galaxy we indicate the identification and technique used to
determine the halo's flattening: 1) Flaring gas layer
\protect\cite{rpO96a,BCV97}; 2) Warping gas layer
\protect\cite{HS94}; 3) X-ray isophotes, \protect\cite{BC98}; 4) Polar
ring galaxies \protect\cite{ACCHS93,SRJF94,SP95}; 5) precessing dusty
disk \protect\cite{SKD92}.  The dotted line shows the predicted
distribution for the shapes of halos in a cold dark matter cosmology
($q=0.5 \pm 0.15$; Dubinski 1994).
}
\end{figure}

In Fig.~\ref{fig:Halo_Shapes_cmp}, we summarize the existing estimates
of halo shape, derived using a variety of techniques.  It is evident
from this figure that there is only a very limited amount of data
available for measuring the distribution of $q$ in galaxies.  Rather
more worrying, though, is the fact that different techniques seem to
yield systematically different answers. The warping gas layer method,
in which the shape of the halo is inferred by treating any warp in a
galaxy's gas layer as a bending mode in a flattened potential
\cite{HS94}, seems to imply that dark halos are close to spherical,
with $q \ga 0.8$.  Conversely, the flaring gas layer technique, which
determines the halo shape by assuming that the gas layer is in
hydrostatic equilibrium in the galaxy's gravitational potential and
uses the thickness of the layer as a diagnostic for the distribution
of mass in the galaxy \cite{rpO96a}, yields much flatter halo shape
estimates with $q \la 0.4$.  Although the numbers involved are rather
small, there do seem to be real differences between the results
obtained by the different methods.  Thus, either these techniques are
being applied to systematically different classes of galaxy, or at
least some of the methods are returning erroneous results.

In order to determine which techniques are credible, we really need to
apply several methods to a single galaxy, to see which produce
consistent answers.  As a first step towards such cross-calibration,
this paper compares the shape of our own galaxy's dark matter halo as
inferred by two distinct techniques:

\begin{enumerate}

\item {\it Stellar kinematics}.  Our position within the Milky Way
means that we have access to information for this galaxy that cannot
be obtained from other systems.  In particular, it is possible to
measure the total column density of material near the Sun using
stellar kinematics.  After subtracting off the other components, we
can infer the local column density of dark matter.  By comparing this
density close to the plane of the Galaxy to the over-all mass as
derived from its rate of rotation, we can obtain a direct measure of
the shape of the halo.

\item {\it The flaring gas layer method}.  As outlined above, this
technique assumes that the H{\sc i} emission in the Milky Way comes from
gas in hydrostatic equilibrium in the Galactic potential, from which the
shape of the dark halo is inferred.  Since the results of previous
applications of this method give results somewhat out of line with the
other techniques, it is important to assess the method's credibility. 

\end{enumerate}

As well as providing a check on the validity of the flaring gas layer
method, these analyses will also provide another useful datum on the
rather sparsely populated Fig.~\ref{fig:Halo_Shapes_cmp}.

The remainder of the paper is laid out as follows.  Both of the above
methods rely on knowledge of the Milky Way's rotation velocity as a
function of Galactic radius -- its ``rotation curve'' -- so
Section~\ref{sec:rot_curve} summarizes the data available for
estimating this quantity, and the dependence of the inferred rotation
curve on the assumed distance to the Galactic centre and local
rotation velocity.  The analysis of the shape of the dark halo
requires that we decompose the Milky Way into its visible and dark
matter components, so Section~\ref{sec:mass_mod} discusses the
construction of a set of models consistent with both the photometric
properties of the Galaxy, and its mass properties as inferred from the
rotation curve.  In Section~\ref{sec:q_from_stars} we show how these mass
models can be combined with the local stellar kinematic measurements
to determine the shape of the dark halo.  Section~\ref{sec:q_from_gas}
presents the application of the gas layer flaring technique to the
mass models.  Section~\ref{sec:q_from_both} combines the results
derived by the two techniques and assesses their consistency.  The
broader conclusions of this work are drawn in
Section~\ref{sec:conclusions}.

\section{The observed rotation curve}
\label{sec:rot_curve} 
Our position within the Milky Way complicates the geometry when
studying its structure and kinematics.  It is therefore significantly
harder to determine our own galaxy's rotation curve, $\Theta(R)$, than
it is to derive those for external systems.  More directly accessible
to observation than $\Theta(R)$ is the related quantity
\begin{equation}
W(R) = R_0\left[{\Theta(R) \over R} - {\Theta_0 \over R_0}\right],
\label{eq:WofR}
\end{equation}
where $R_0$ and $\Theta_0$ are the distance to the Galactic centre and
the local circular speed.  If one assumes that material in the galaxy
is in purely circular motion, then some simple geometry shows that, for
an object at Galactic coordinates $\{l,b\}$ with a line-of-sight
velocity $v_{\rm los}$,
\begin{equation}
W = {v_{\rm los} \over \sin l \cos b}
\label{eq:Wdef}
\end{equation}
\noindent (Binney \& Merrifield 1998, \S9.2.3).  Thus, if one measures
the line-of-sight velocities for a series of objects at some radius
$R$ in the Galaxy, one has an estimate for $W(R)$.  By adopting values
for $R_0$ and $\Theta_0$, one can then use equation~(\ref{eq:WofR}) to
determine the rotation speed at that radius.

In practice, the difficulty lies in knowing the Galactic radii of the
objects one is looking at.  One solution is to look at standard
candles, whose distances can be estimated, and hence whose radii in
the Galaxy can be geometrically derived.  Alternatively, one can
select the subset of some tracer -- usually H{\sc i} or H$_2$ gas -- whose
line-of-sight velocities and Galactic coordinates places it at the
same value of $W$, and hence at the same radius.  One can then use the
properties of this cylindrical slice through the Galaxy to infer its
radius.  For example, in the inner Galaxy, all the material in a ring
of radius $R$ will lie at Galactic longitudes of less than $l_{\rm
max} = \sin^{-1} (R/R_0)$, so one can use the extent of the
emission on the sky of each $W$-slice to infer its radius -- an
approach traditionally termed the ``tangent point method'' [see, for
example, Malhotra (1994,1995)].  At radii greater than $R_0$ this
method is no longer applicable as the emission will be visible at all
values of $l$.  However, by assuming that the thickness of the gas
layer does not vary with azimuth, one can use the observed variation
in the angular thickness of the layer with Galactic longitude to
estimate the radii of such $W$ slices (Merrifield 1992).

For the remainder of this paper, we use Malhotra's (1994, 1995)
tangent point analysis to estimate $W(R)$ in the inner Galaxy.  For
the outer Galaxy, we have combined Merrifield's (1992) data with Brand
\& Blitz's (1993) analysis of the distances to H{\sc II} regions, from which
standard candle analysis the rotation curve can be derived.  

In order to convert $W(R)$ into $\Theta(R)$ using
equation~(\ref{eq:WofR}), we must adopt values for the Galactic
constants, $R_0$ and $\Theta_0$.  Unfortunately, there are still
significant uncertainties in these basic parameters.  In the case of
$R_0$, for example, the extensive review by Reid (1993) discussed
measurements varying between $R_0 = 6.9 \pm 0.6\,{\rm kpc}$ and $R_0 =
8.4 \pm 0.4\,{\rm kpc}$.  Even more recently there have been few signs
of convergence: Layden et al.\ (1996) used RR Lyrae stars as standard
candles to derive $R_0 = 7.2 \pm 0.7\,{\rm kpc}$, while Feast \&
Whitelock (1997) used a Cepheid calibration to obtains $R_0 = 8.5 \pm 0.5\,{\rm
kpc}$.  The constraints on $\Theta_0$ are similarly weak: a recent
review by Sackett (1997) concluded that a value somewhere in the range
$\Theta_0 = 210 \pm 25\,{\rm km}\,{\rm s}^{-1}$ provided the best
current estimate.  It should also be borne in mind that the best
estimates for $R_0$ and $\Theta_0$ are not independent.  Analysis of the
local stellar kinematics via the Oort constants gives quite a
well-constrained value for the ratio $\Theta_0/R_0 = 26.4 \pm 1.9\,{\rm
km}\,{\rm s}^{-1}\,{\rm kpc}^{-1}$ (Kerr \& Lynden-Bell 1986), so a
lower-than-average value of $R_0$ is likely to be accompanied by a
lower-than-average value for $\Theta_0$.  Currently, the best available
measure of the local angular velocity of the Milky Way is based on VLBI
proper motion measurements of SgrA$^*$.  Assuming that SgrA$^*$ is at
rest with respect to the Galactic centre, Reid et al.\ (1999) find
$\Theta_0/R_0 = 27.25 \pm 2.5 {\rm km}\,{\rm s}^{-1}\,{\rm kpc}^{-1}$,
consistent with the value proposed by Kerr \& Lynden-Bell.

\begin{figure}
 \epsffile{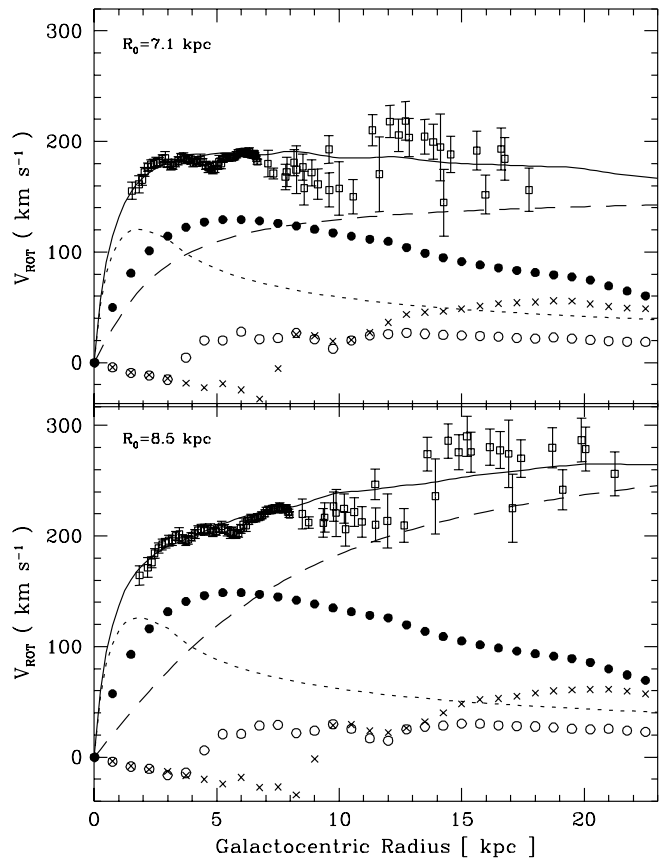} 
\caption{
\label{fig:MW_Rotation_Curves} The rotation curve for the Milky Way
for values of $R_0 = 7.1\,{\rm kpc}$, $\Theta_0 = 185\,{\rm km}\,{\rm
s}^{-1}$, and $R_0 = 8.5\,{\rm kpc}$, $\Theta_0 = 220\,{\rm km}\,{\rm
s}^{-1}$.  The figure also shows one of the ways in which the rotation
curve can be decomposed into the contributions from different mass
components: the bulge (dotted line); the stellar disk (filled
circles); the H{\sc i} layer (crosses, where negative values mean that
the force is directed outwards); the $H_2$ layer (circles); and the
dark halo (dashed line).  The best fit model, which is obtained by
summing the individual components in quadrature, is shown as a full
line.  
}

\end{figure}

To illustrate the effect of the adopted values of $\Theta_0$ and $R_0$
on the derived rotation curve, Fig.~\ref{fig:MW_Rotation_Curves} shows
$\Theta(R)$ for two of the more extreme plausible sets of Galactic
parameters.  Clearly, the choice of values for these quantities
affects such basic issues as whether the rotation curve is rising or
falling in the outer Galaxy.  Given the current uncertainties, it
makes little sense to pick fixed values for the Galactic constants,
so in the following analysis we consider models across a broad
range of values, $5.5\,{\rm kpc} < R_0 < 9\,{\rm kpc}$, $165\,{\rm
km}\,{\rm s}^{-1} < \Theta_0 < 235\,{\rm km}\,{\rm s}^{-1}$.

\section{Mass models} 
\label{sec:mass_mod}

In order to relate the rotation curve to the shape of the dark halo, we
must break the gravitational potential responsible for the observed
$\Theta(R)$ into the contributions from the different mass components. 
As is usually done in this decomposition
\cite{egvAS86,smK87,kBeg89,LF89,ahB92,rpO95,rpO96b,wDjB98}, we adopt a
model consisting of a set of basic components:

\begin{enumerate}

\item {\it A stellar bulge}.  Following Kent's (1992) analysis of the
Galaxy's K-band light distribution, we model the Milky Way's bulge by a
``boxy'' density distribution,

\begin{equation}
\rho_b(R,z) \propto K_0(s/h_b),\ \hbox{where}\ s^4 = R^4 + (z/q_b)^4.
\end{equation}

\noindent This modified Bessel function produces a bulge that appears
exponential in projection.  The observed flattening of the K-band light
yields $q_b = 0.61$, and its characteristic scalelength is $h_b =
670\,{\rm pc}$ (Kent 1992).  The constant of proportionality depends on
the bulge mass-to-light ratio, $\Upsilon_b$, which we leave as a free
parameter. 

\item {\it A stellar disk}.  The disk is modelled by a density
distribution,
\begin{equation}
\rho_d(R,z) \propto \exp(-R/h_d) {\rm sech}(z/2z_d).
\end{equation}
\noindent The first term gives the customary radially-exponential
disk.  The appropriate value for the scalelength, $h_d$, is still
somewhat uncertain -- Kent, Dame \& Fazio (1991) estimated $h_d = 3
\pm 0.5\,{\rm kpc}$, while Freudenreich (1998) found a value of
$2.5\,{\rm kpc}$.  We therefore leave this parameter free to vary
within the range $2\,{\rm kpc} \le h_d \le 3\,{\rm kpc}$.  The
$z$-dependence adopts van der Kruit's (1988) compromise between a
${\rm sech}^2$ isothermal sheet and a pure exponential.  For
simplicity, we fix the scaleheight at $z_d = 300\,{\rm pc}$.  However,
the exact $z$-dependence of $\rho_d$ was found to have no impact on
any of the following analysis.  Once again, the constant of
proportionality depends on the mass-to-light ratio of the disk,
$\Upsilon_d$, which we leave as a free parameter.
\item {\it A gas disk}.  From the H{\sc i} data given by Burton (1988)
and Malhotra (1995) and the H$_2$ column densities from Bronfman et
al.\ (1988) and Wouterloot et al.\ (1990), we have inferred the
density of gas as a function of radius in the Galaxy.  This analysis
treats the gas as an axisymmetric distribution, and neglects the
contribution from ionized phases of the interstellar medium, but does
include a 24\% contribution by mass from helium (Olive \& Steigman
1995).

\item {\it A dark matter halo}.  We model the dark halo as a flattened
non-singular isothermal sphere, which has a density distribution

\begin{equation}
\rho_h(R,z) = \rho_h {R_h^2 \over R_h^2 + R^2 + (z/q)^2},
\end{equation}

\noindent where $\rho_h$ is the central density, $R_h$ is the halo core
radius, and $q$ is the halo flattening, which is the key parameter in
this paper.  \end{enumerate}

The procedure for calculating a mass model from these components is
quite straightforward.  For each pair of Galactic constants, $R_0$ and
$\Theta_0$, we vary the unknown parameters $\{\Upsilon_b, h_d,
\Upsilon_d, \rho_h, R_h, q\}$ to produce the mass model that has a
gravitational potential, $\Phi(R, z)$, such that \begin{equation} v(R) =
\left(R {\partial\Phi\over\partial R}\biggr|_{z=0}\right)^{1/2}
\end{equation} provides the best fit (in a minimum $\chi^2$ sense) to
the observed rotation curve, $\Theta(R)$.  Examples of two such best-fit
models are shown in Fig.~\ref{fig:MW_Rotation_Curves}. 

As is well known
\cite{egvAS86,smK87,kBeg89,LF89,ahB92,rpO95,rpO96b,wDjB98}, such mass
decompositions are by no means unique: there is near degeneracy between
the various unknown parameters, so many different combinations of the
individual components can reproduce the observed rotation curve with
almost equal quality of fit.  We have therefore searched the entire
$\{\Upsilon_b, h_d, \Upsilon_d, \rho_h, R_h, q\}$ parameter space to
find the complete subset of values that produce fits in which the
derived value of $\chi^2$ exceeded the minimum value by less than unity.

\section{Halo flattening from local stellar kinematics}
\label{sec:q_from_stars} 

Although the analysis of the previous section tells us something about
the possible range of mass models for the Milky Way, it does not place
any useful constraint on the shape of the dark halo: for any of the
adopted values of $R_0$ and $\Theta_0$, one can find acceptable mass
models with highly-flattened dark halos ($q \sim 0$), round dark halos
($q \sim 1$), and even prolate dark halos ($q > 1$).  We therefore
need some further factor to discriminate between these models.

One such constraint is provided by stellar kinematics in the solar
neighbourhood.  Studies of the motions of stars in the Galactic disk
near the Sun imply that the total amount of mass within 1.1 kpc of the
Galactic plane is $\Sigma_{1.1} = (71 \pm 6) M_{\odot}\,{\rm pc}^{-2}$
(Kuijken \& Gilmore 1991).  Clearly, the value of $\Sigma_{1.1}$
provides an important clue to the shape of the Milky Way's dark halo:
in general, a model with a highly-flattened dark halo will place a lot
of mass near the Galactic plane leading to a high predicted value for
$\Sigma_{1.1}$, while a round halo will distribute more of the dark
matter further from the plane, depressing $\Sigma_{1.1}$.

The dark halo is not the only contributor to $\Sigma_{1.1}$.  In
particular, the stellar disk has a surface density in the solar
neighbourhood, $\Sigma_*$, which may contribute a significant fraction
of $\Sigma_{1.1}$.  There must therefore be a fairly simple relation
between the adopted value of $\Sigma_*$ and the inferred halo
flattening, $q$.  Specifically, as one considers larger possible
values of $\Sigma_*$, the amount of dark matter near the plane must
decrease so as to preserve the observed value of $\Sigma_{1.1}$.  Such
a decrease can be achieved by increasing $q$, thus making the dark
halo rounder.

\begin{figure}
\epsffile{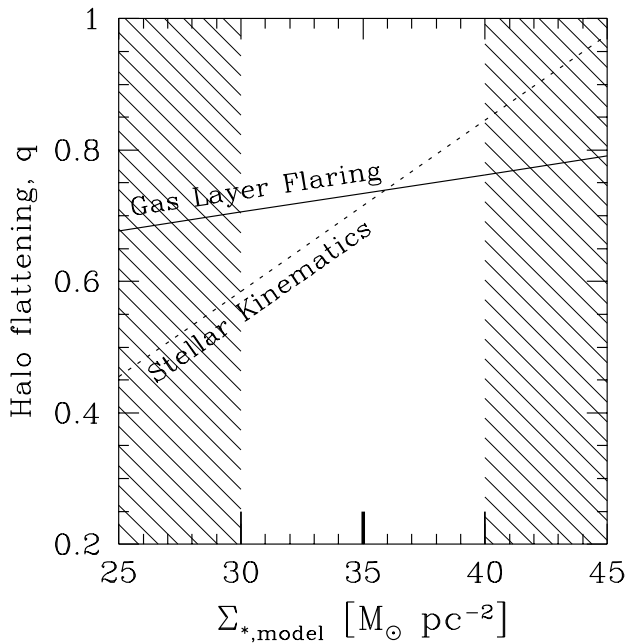} \caption{ \label{fig:q_Sigma_star}
The flattening of the model Galactic halo, $q$, as a function of the
adopted model's column of stellar disk mass in the Solar
neighbourhood, $\Sigma_*$.  We show the best fits to the stellar
kinematic constraints (dotted line) and the flaring of the H{\sc i}
layer (full line).  The shaded region shows values of $\Sigma_*$ that
lie more than 1-$\sigma$ from the observed value.  This particular set
of calculations has been made assuming $R_0 = 7.1\,{\rm kpc}$ and
$\Theta_0=185{\rm\,km \,s}^{-1}$.}
\end{figure}

This inter-relation is illustrated in Fig.~\ref{fig:q_Sigma_star}.
Here, for one of the sets of possible Galactic constants, we have
considered all the mass models that produce an acceptable fit to the
rotation curve and reproduce the meaured value of $\Sigma_{1.1}$.  For
each acceptable model, we have extracted the value of the halo
flattening $q$ and the mass of the stellar disk in the solar
neighbourhood, $\Sigma_*$.  For the reasons described above, these
quantities are tightly correlated, and Fig.~\ref{fig:q_Sigma_star}
shows the linear regression between the two.

Clearly, we have not yet derived a unique value for the halo
flattening: by selecting models with different values of $\Sigma_*$,
we can still tune $q$ to essentially any value we want.  However,
$\Sigma_*$ is not an entirely free parameter.  From star-count
analysis, it is possible to perform a stellar mass census in the solar
neighbourhood, and thus determine the local stellar column density.
Until relatively recently, this analysis has been subject to
significant uncertainties, with estimates as high as $\Sigma_* =
145M_{\odot}\,{\rm pc}^{-2}$ \cite{jnB84b}.  However, there is now
reasonable agreement between the various analyses, with more recent
published values of $35 \pm 5M_{\odot}\,{\rm pc}^{-2}$ (Kuijken \&
Gilmore 1989) and $26 \pm 4M_{\odot}\,{\rm pc}^{-2}$ (Gould, Bahcall
\& Flynn 1997).  If we adopt Kuijken \& Gilmore's slightly more
conservative error bounds on $\Sigma_*$, the shaded regions in
Fig.~\ref{fig:q_Sigma_star} no longer represent acceptable models as
they predict the wrong local disk mass density, so we end up with a
moderately well-constrained estimate for the halo flattening of $q =
0.7 \pm 0.1$.

Note, though, that although this analysis returns a good estimate for
$q$, Fig.~\ref{fig:q_Sigma_star} only shows the value obtained for a
particular set of Galactic constants.  As discussed above, changing
the adopted values for $R_0$ and $\Theta_0$ alters the range of
acceptable mass models, which, in turn, will alter the derived
correlation between $\Sigma_*$ and $q$.  The uncertainties in the
Galactic constants are still sufficiently large that the absolute
constraints on $q$ remain weak.  Nevertheless, it is to be hoped that
the measurements of $R_0$ and $\Theta_0$ will continue to improve over
time, leading to a unique determination of $q$ by this method.
Further, as we shall see below, it may be possible to attack the
problem from the other direction by using other estimates of $q$ to
help determine the values of the Galactic constants.

\section{Halo flattening from gas layer flaring}
\label{sec:q_from_gas}

We now turn to the technique developed by Olling (1995) for measuring
the shape of a dark halo from the observed thickness of a galaxy's gas
layer.  In essence the approach is similar to the stellar-kinematic
method described above: the over-all mass distribution of the Milky
Way is inferred from its rotation curve, while the degree to which
this mass distribution is flattened is derived using the properties of
a tracer population close to the Galactic plane.  In this case, the
tracer is provided by the H{\sc i} emission from the Galactic gas
layer.  The thickness of the Milky Way's gas layer is dictated by the
hydrostatic balance between the pull of gravity toward the Galactic
plane and the pressure forces acting on the gas.  As the density of
material in the Galaxy decreases with radius, the gravitational force
toward the plane becomes weaker, so the equilibrium thickness of the
layer becomes larger, giving the gas distribution a characteristic
``flared'' appearance.  The exact form of this flaring depends on the
amount of mass close to the plane of the Galaxy.  Thus, by comparing
the observed flaring to the predictions of the hydrostatic equilibrium
arrangement of gas in the mass models developed in
Section~\ref{sec:mass_mod}, we can see what degree of halo flattening
is consistent with the observations.

\begin{figure*}
 \epsffile{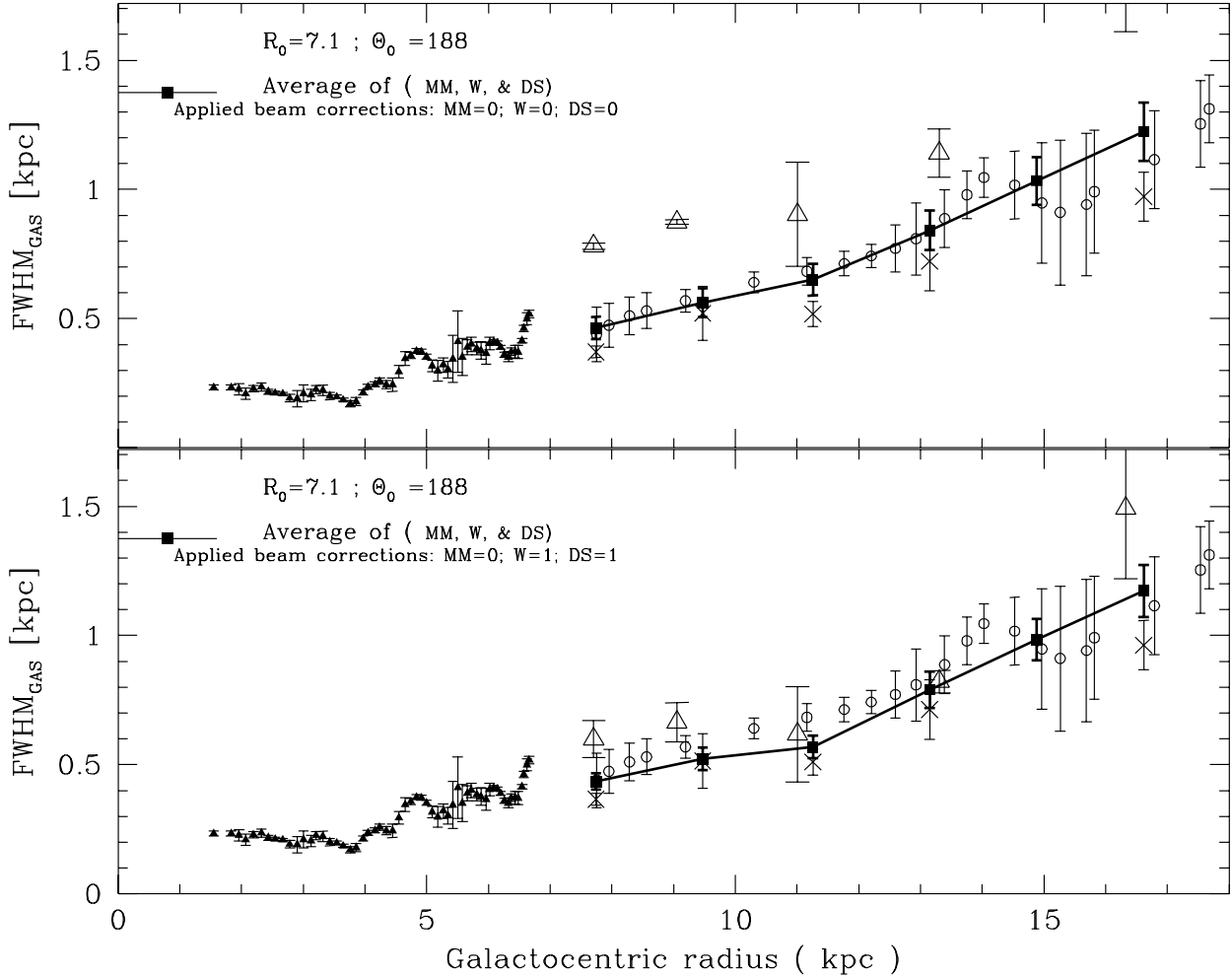}
 \caption{ \label{fig:Observed_Flaring} The thickness of the
H{\sc i} layer of the Milky Way.  The widths in the inner Galaxy were
taken from Malhotra (1995).  For the outer Galaxy, we plot the widths
from Diplas \& Savage (1991; open triangles), Wouterloot et al.\
(1992; crosses), Merrifield (1992; open circles), and the average (thick
full line \& filled squares).  The top panel represents the ``raw''
measurements, in bottom panel we present the beam-size corrected widths. 
} 

\end{figure*}

\subsection{The observed flaring of the gas layer} Before we can apply
this technique, we need to summarize the observational data available
on the flaring of the Galactic H{\sc i} layer.  Merrifield (1992)
calculated the thickness of the gas layer across a wide range of
Galactic azimuths in his determination of the outer Galaxy rotation curve.
As a check on the validity of that analysis, we have also drawn on the
work of Diplas \& Savage (1991) and Wouterloot et al.\ (1992), which
derived the gas layer thicknesses across a more limited range of
azimuths.  For completeness, we have also included the data for the
inner Galaxy as derived by Malhotra (1995).  The resulting values for
the full-width at half maximum (FWHM) of the density of gas
perpendicular to the plane are given in the upper panel in
Fig.~\ref{fig:Observed_Flaring}.  Note that, once again, the results
depend on the adopted values of the Galactic constants -- since the
radii in the Galaxy of the various gas elements were derived from
their line-of-sight velocities via equations~(\ref{eq:WofR}) and
(\ref{eq:Wdef}), the values of $R$ in Fig.~\ref{fig:Observed_Flaring}
depend on $R_0$ and $\Theta_0$\footnote{Merrifield's method for
determining the thickness of the gas layer also exposes some of the
shortcomings of more traditional methods.  If the wrong values for
$R_0, \Theta_0$ and $\Theta(R)$ are chosen, the inferred thickness of
the gas layer (at a given $R$) will show a systematic variation with
Galactocentric azimuth.  It is possible to correct published data for
this effect, but only if the assumed rotation curve, Galactic
constants, as well as the thickness of the gas layer as a function of
azimuth are specified.}.

It would appear from this figure that there are some discrepancies
between the various measurements in the outer Galaxy.  After some
investigation, we established that these differences can be attributed
to the effects of the beam sizes of the radio telescopes with which
the observations were made.  Such resolution effects will mean that
the FWHM of the gas will tend to be overestimated.  In the lower
panel, we show what happens when the appropriate beam correction is
made to the data from Diplas \& Savage (1991) and Wouterloot et al.\
(1992).  No similar correction is required for the Merrifield (1992)
analysis, as in that work the derived value of the FWHM was dominated
by gas towards the Galactic anti-centre, which lies at relatively small
distances from the Sun, and so the beam correction is small.  Clearly,
this correction brings the various data sets into much closer agreement.
We therefore adopt the mean curve shown in this panel for the
following analysis; the error bars shown represent the standard error
obtained on averaging the various determinations.

\subsection{Sources of pressure support}
In order to compare the observed gas layer flaring to the predictions
for a gas layer in hydrostatic equilibrium, we must address the source
of the pressure term in the hydrostatic equilibrium equation.  The
most obvious candidate for supporting the H{\sc i} layer comes from
its turbulent motions.  In the inner Galaxy, the H{\sc i} is
observed to have a velocity dispersion of $\sigma_g = 9.2{\rm\,km
\,s}^{-1}$ independent of radius \cite{sM95}, and we assume that this
value characterizes the turbulent motions of the gas throughout the
Galaxy, providing a kinetic pressure term.

Potentially, there may be other forces helping to support the Galactic
H{\sc i} layer: non-thermal pressure gradients associated with magnetic
fields and cosmic rays may also provide a net force to resist the pull
of gravity on the Galactic H{\sc i}.  However, the analysis we are doing
depends most on the properties of the H{\sc i} layer at large radii in the
Milky Way, where star formation is almost non-existent, so energy
input into cosmic rays and magnetic fields from stellar processes is
likely to be unimportant.

Our concentration on the properties of gas at large radii also
eliminates another potential complexity.  In the inner Galaxy, the
interstellar medium comprises a complicated multi-phase mixture of
molecular, atomic and ionized material.  A full treatment of the
hydrostatic equilibrium of such a medium is complicated, as gas can
transform from one component to another, so all components would have
to be considered when calculating hydrostatic equilibrium.  For the
purpose of this paper, it is fortunate that at the low pressures
characteristic of the outer Galaxy it is not possible to maintain both
the cold molecular phase and the warm atomic phase
\cite{pM93,WHKTB95}.  Braun \& Walterbos (1992) and Braun (1997,1998)
have shown that the cold phase disappears when the B-band surface
brightness of a galaxy falls below the 25th magnitude per square
arcsecond level, which occurs at $\sim 1.5 R_0$ in the Milky Way
(Binney \& Merrifield 1998, \S10.1).  Further, the ionized fraction of
the ISM is expected to decrease with distance as it is closely
associated with sites of star formation \cite{FWGH96,WHL97}.
Ultimately, the ionizing effects of the extragalactic background
radiation field become significant, but only when the H{\sc i} column
density falls below about $1 M_{\odot}\,{\rm pc}^{-2}$
\cite{pM93,DS94,rpO96b}, which lies well beyond the radii we are
considering here.

For the current analysis, it therefore seems reasonable to treat the
Milky Way's gas layer as a single isothermal component supported
purely by its turbulent motions.  Moreover, this assumption has been
made in the previous implementations of the gas flaring method
\cite{rpO96a,BCV97}.  A principal objective of this paper is to test
the validity of those analyses by comparing results obtained by the
flaring technique to those obtained by other methods.  It is therefore
important that we make the same assumption of a single isothermal
component in the present study.

\subsection{Fitting to model gas layer flaring}
We are now in a position to compare observation and theory.  The
technique for calculating the gas layer thickness in different mass
models has been described in detail by Olling (1995).  In brief, for
each model, at each radius $R$, one integrates the hydrostatic
equilibrium equation,
\begin{equation}
{\partial\Phi \over \partial z} 
            = -{1 \over \rho_g} {\partial \rho_g \sigma_g^2 \over \partial z},
\end{equation}
to obtain the gas density distribution, $\rho_g(R, z)$.  The FWHM of
this model gas distribution can then be compared directly with the
observations.

\begin{figure}
 \epsffile{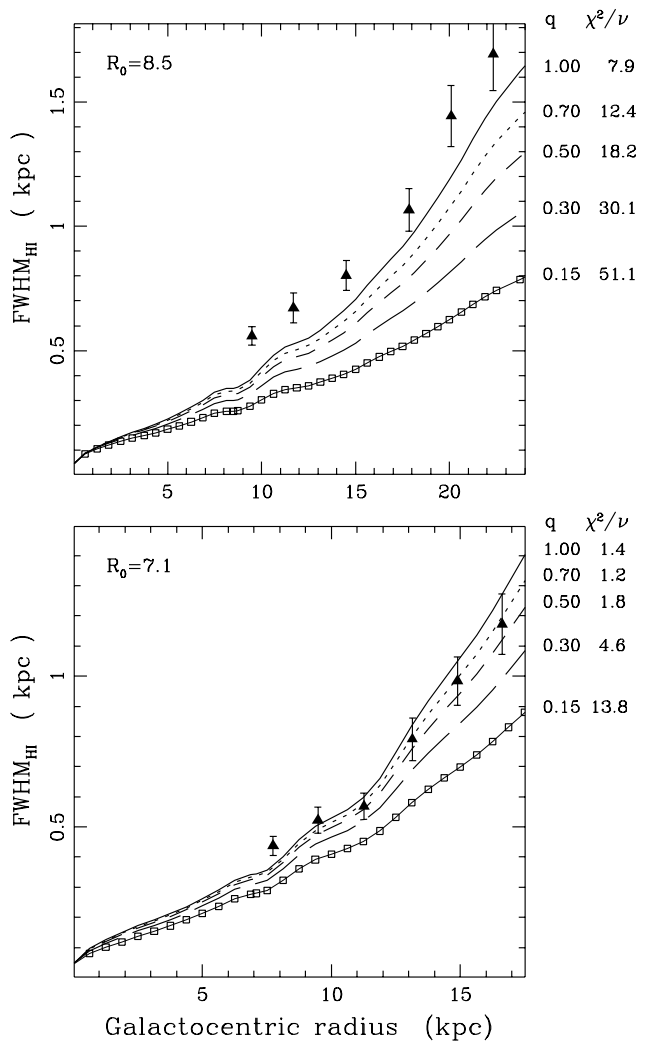} 
\caption{\label{fig:Obs_Mod_Flaring} The observed flaring of the
Galaxy's gas layer (triangles) compared with the flaring predicted for
a number of mass models that reproduce the observed rotation curve and
value of $\Sigma_*$.  The models in the top panel have been calculated
assuming $R_0 = 8.5\,{\rm kpc}$ and a disk mass-to-light ratio
$\Upsilon_{\rm d, K-band} = 0.60$; the bottom panel shows models with
$R_0 = 7.1\,{\rm kpc}$ and $\Upsilon_{\rm d,K-band} = 0.41$.  The
models shown here have a disk scale-length of $h_d = 2.5\,{\rm kpc}$.
The halo flattening of the model and the reduced $\chi^2$ value of the fit
is shown for each case.}
\end{figure}

The results of such calculations are illustrated by the examples shown
in Fig.~\ref{fig:Obs_Mod_Flaring}.  The basic trends in this analysis are
clearly demonstrated by these examples.  The decrease in total density
with radius leads to a dramatic flaring in the model-predicted gas layer
thickness, just as is seen in the observations.  For a flatter model
dark halo, the mass is more concentrated toward the plane of the Galaxy,
squeezing the H{\sc i} layer into a thinner distribution. 

Once again, the results depend quite sensitively on the choice of
Galactic constants, since these values affect both the gas
distribution as inferred from observations and the acceptable mass
models as inferred from the rotation curve.  As is apparent from
Fig.~\ref{fig:Obs_Mod_Flaring}, none of the $R_0 = 8.5\,{\rm kpc}$
models fits the observations.  In fact, to match the observed layer
width one would require a substantially prolate dark matter halo with
$q \sim 1.5$, which none of the current dark matter scenarios predict.
For $R_0 = 7.1\,{\rm kpc}$, on the other hand, a very good fit is
obtained for models with a halo flatness of $q \sim 0.7$.  Such models
even reproduce the observed inflection in the variation of the gas
layer width with radius at $R \sim 10\,{\rm kpc}$.

For each plausible set of Galactic constants, we can carry out a
similar analysis to that in Section~\ref{sec:q_from_stars} to see what
range of values of $q$ are consistent with the observed gas layer
flaring.  Because this technique relies on data from large radii in
the Galaxy, where the dark matter halo is the dominant source of mass,
the flaring predicted by the models depends very little on the
properties of the disk and bulge.  Unlike the stellar kinematic
analysis, therefore, one cannot trade off the mass in the disk against
the mass near the plane from a more-flattened dark halo.  This
difference is illustrated in Fig.~\ref{fig:q_Sigma_star}, which shows
the way that the value of $q$ inferred from the gas layer flaring
depends on the properties of the stellar disk (as parameterized by the
model's column density of stars in the Solar neighbourhood).  As for
the stellar-kinematic analysis, there is a well-defined correlation
between $q$ and $\Sigma_*$, but, for the reasons described above, the
trend for the current method is very much weaker, and, within the
observationally-allowed range for $\Sigma_*$, $q$ is tightly
constrained to $0.73 \pm 0.03$.

Although this constraint is remarkably good, it should be borne in
mind that it is still dependent on the adopted values for the Galactic
constants.  As we have seen above, larger values for $R_0$ lead to
rounder, or even prolate estimates for halo shape, so we will not
obtain an unequivocal measure for $q$ from this method until the
Galactic constants are measured more accurately

\section{Combining the techniques}
\label{sec:q_from_both}

The different slopes of the two relations in
Fig.~\ref{fig:q_Sigma_star} raises an interesting possibility.
Clearly, for a consistent picture, one must use a single mass model to
reproduce both the stellar-kinematic constraint on the mass in the
solar neighborhood and the observed flaring of the gas layer.  Thus,
although there are whole linear loci in this figure of models with
different values of $q$ and $\Sigma_*$ that satisfy each of these
constraints individually, there is only the single point of
intersection between these two lines where the model fits both the
stellar-kinematic constraint and the observed flaring of the gas
layer.  Hence, for given values of $R_0$ and $\Theta_0$, one predicts
unique values for $q$ and $\Sigma_*$.

We have therefore repeated the analysis summarized in
Fig.~\ref{fig:q_Sigma_star} spanning the full range of plausible values
for $R_0$ and $\Theta_0$, and calculated the mutually-consistent
estimates for $\Sigma_*$ and $q$ for each case.  In order to reduce the
computational complexity of this large set of calculations to manageable
proportions, we made use of Olling's (1995) fitting formula,
which showed that, if self-gravity is negligible, one can approximate
the model-predicted thickness of the gas layer by the relation
\begin{equation}
{\rm FWHM}(R) \approx \sqrt{13.5q \over 1.4 + q} {\sigma_g \over v_{h, \infty}}
                                                          \sqrt{R_h^2 + R^2},
\label{eq:FWHMapprox}
\end{equation}
\noindent where $v_{h, \infty}$ is the circular rotation speed of the
dark halo component at large radii.  Allowing for the self-gravity of
the gas layer, one obtains a similar formula.  Comparing the values
derived from this formula to sample results obtained by the full
integration process described above, we found that the approximation
based on equation~(\ref{eq:FWHMapprox}) matches the detailed
integration for $R > 1.75R_0$.  We therefore only used the data from
beyond this radius in the following analysis, in which the model gas
layer thickness was estimated using the approximate formula.

\begin{figure*}
\epsffile{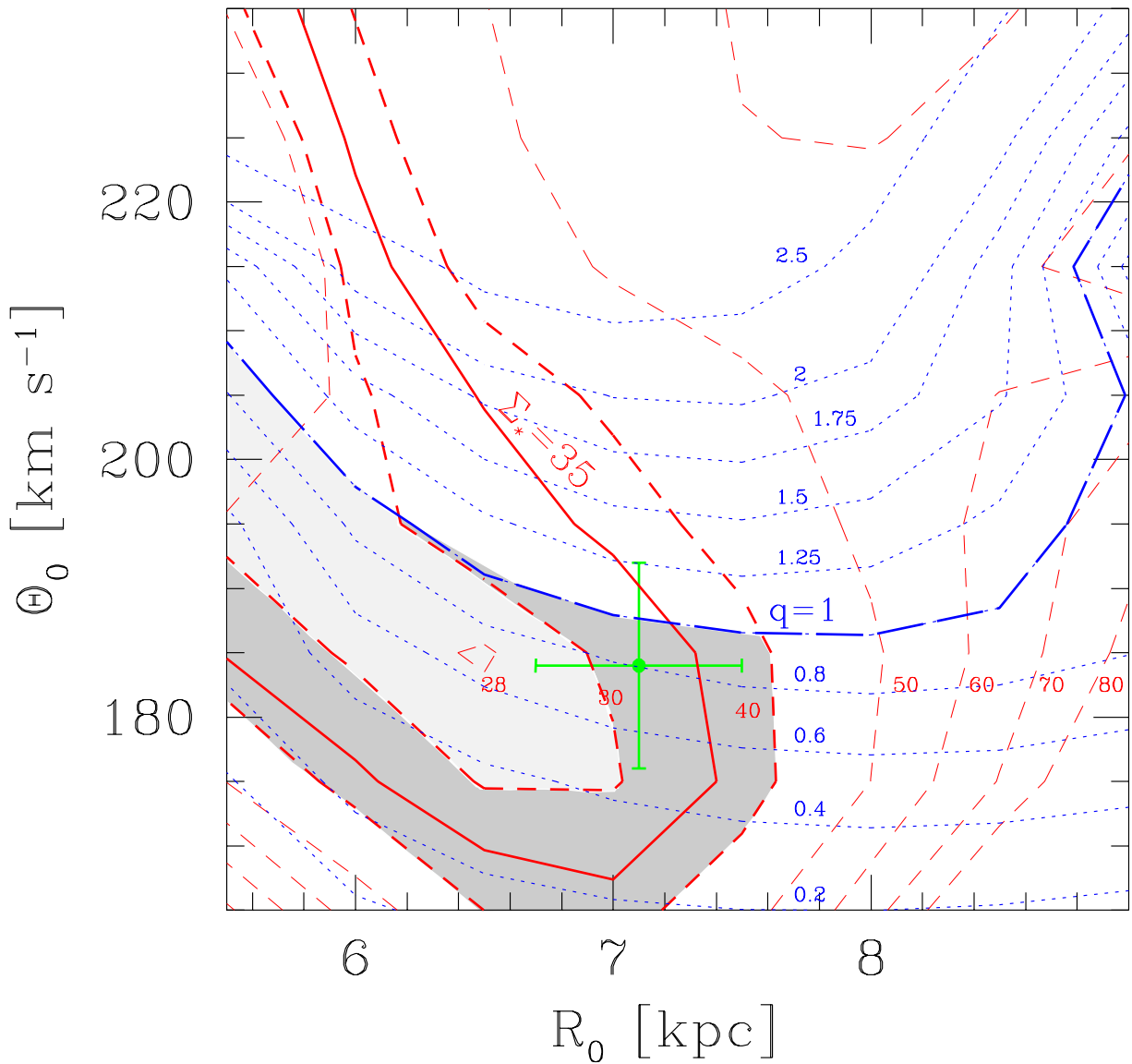} \caption{ \label{fig:Sstr_q_R0_T0}
Contours of the stellar column density in the Solar neighbourhood
($\Sigma_*$; long dashed lines) and halo flattening ($q$; dotted
lines) as a function of the adopted values for the Galactic constants,
$R_0$ and $\Theta_0$.  The heavy full line and the heavy dashed line
corresponds to Kuijken \& Gilmore's (1989) determination of
$\Sigma_*$, and the $\pm1-\sigma$ values.  The heavily-shaded region
corresponds to parts of parameter space consistent with these values
of $\Sigma_*$ that produce an oblate halo.  The lightly-shaded area
gives the corresponding region if we adopt Gould, Bahcall and Flynn's
(1997) values for $\Sigma_*$.  The cross shows the determination of
the Galactic constants based on an analysis of the Oort constants
(Olling \& Merrifield 1998).  }
\end{figure*}

The values obtained for $q$ and $\Sigma_*$ for each possible pair of
Galactic constants are presented in Fig.~\ref{fig:Sstr_q_R0_T0}.
Thus, for example, for values of $R_0 = 7.1\,{\rm kpc}$ and $\Theta_0
= 185\,{\rm km}\,{\rm s}^{-1}$, one finds $\Sigma_* = 35
M_{\odot}\,{\rm pc}^{-2}$ and $q = 0.7$, corresponding to the
intercept that we previously calculated in
Fig.~\ref{fig:q_Sigma_star}.

Figure~\ref{fig:Sstr_q_R0_T0} places some interesting limits on the
properties of the Milky Way.  For example, if we maintain our
prejudice that the halo should be oblate ($q < 1$), then, unless we
adopt a particularly extreme value for $R_0$, we find that $\Theta_0$
must be less than $\sim 190\,{\rm km}\,{\rm s}^{-1}$.  If we also
adopt Kuijken \& Gilmore's (1989) measurement of $\Sigma_* = 35 \pm
5M_{\odot}\,{\rm pc}^{-2}$, we find that only models within the
heavily-shaded region of Fig.~\ref{fig:Sstr_q_R0_T0} are acceptable,
placing an upper limit on $R_0$ of $\sim 7.6\,{\rm kpc}$.
Conversely, if one forces the Galactic constants to the IAU standard
values of $R_0 = 8.5\,{\rm kpc}$ and $\Theta_0 = 220\,{\rm km}\,{\rm
s}^{-1}$ (Kerr \& Lynden-Bell 1987), one finds barely-credible values
of $\Sigma_* \sim 60 M_{\odot}\,{\rm pc}^{-2}$ and $q \sim 1.5$.

\section{Conclusions}
\label{sec:conclusions}

The prime objective of this paper has been to check the validity of
techniques for measuring halo flattening by asking whether two
different techniques return consistent values when applied to the
Milky Way.  As we have seen in the last section, the answer is a
qualified ``yes.''  The qualification is that consistency with the
measured stellar column density requires values of the Galactic
constants that differ from those conventionally adopted.  However, as
was discussed in Section~\ref{sec:rot_curve}, the true values of these
constants remain elusive, with estimates spanning the ranges $ 7\,{\rm
kpc} < R_0 < 8.5\,{\rm kpc}$ and $185\,{\rm km}\,{\rm s}^{-1} <
\Theta_0 < 235\,{\rm km}\,{\rm s}^{-1}$.  With such gross
uncertainties, it quite straightforward to pick values that produce an
entirely self-consistent picture.  To underline this point, we have
included on Fig.~\ref{fig:Sstr_q_R0_T0} the results of Olling \&
Merrifield's (1997) estimates for $R_0$ and $\Theta_0$ derived from an
analysis of the Oort constants.  If that analysis is valid, then we
have a consistent model for the Milky Way in which the dark halo has a
flattening of $q \sim 0.8$.  

Ultimately, this analysis will allow us to come to one of two
conclusions:
\begin{enumerate}
\item If future studies confirm low values for $R_0$ and $\Theta_0$
similar to those derived by Olling \& Merrifield (1997), then the
consistency of the two analyses for calculating $q$ imply that the gas
layer flaring technique is valid, adding confidence to the previous
determinations by this method.
\item Conversely, if we learn in future that $R_0$ and $\Theta_0$ are
closer to the more conventional larger values, then the implied values
of $\Sigma_*$ are so far from the observed estimates that one has to
suspect that at least one of the techniques for measuring $q$ is
compromised.  In this case, one would have to look more closely at
some of the assumptions that went into the analysis.  For example,
perhaps the non-thermal pressure forces from cosmic rays and magnetic
fields have a role to play even at large radii in galactic disks.
Alternatively, perhaps the H{\sc i} layer is not close enough to
equilibrium for the hydrostatic analysis to be valid.  Finally, our
assumption of azimuthal symmetry may be invalid.  Strong departures
from axisymmetry could mean that our determination of the thickness
and column density of the gas is compromised, and that the locally
determined values of $\Sigma_{1.1}$ and $\Sigma_*$ may not be
representative for the Galactocentric radius of the Sun.
\end{enumerate}
Assuming for the moment that the analysis is valid, we have another
datum to add to Fig.~\ref{fig:Halo_Shapes_cmp}.  Since the two
previous flaring analyses returned systematically rather small values
of $q \sim 0.3$, it is reassuring that the Milky Way seems to indicate
a larger value of $\sim 0.8$ -- it  appears that the low values
are simply a coincidence arising from the very small number
statistics.  This larger value is inconsistent with the very flat
halos that are predicted by models in which the dark matter consists
of either decaying neutrinos (Sciama 1990) or cold molecular hydrogen
(Pfenniger, Combes \& Martinet 1994).

With the addition of the Milky Way to the data presented in
Fig.~\ref{fig:Halo_Shapes_cmp}, the only technique that stands out as
giving systematically different values for $q$ is the bending mode
analysis of warped gas layers.  In this regard, it is notable that
simulations cast some doubt on the validity of such analyses.  The
method assumes that warps are manifestations of persistent bending
modes that occur when the flattening of a disk and the surrounding
dark halo are misaligned.  However, the simulations show that
gravitational interactions between a misaligned disk and halo rapidly
bring the two back into alignment, effectively suppressing this
mechanism (e.g.\ Dubinski \& Kuijken 1995, Nelson \& Tremaine 1995,
Binney, Jiang \& Dutta 1998).  The number of measurements is still
rather small, but it is at least interesting that if one neglects the
warped gas layer results, the remaining data appear entirely
consistent with the dotted line in Fig.~\ref{fig:Halo_Shapes_cmp},
which shows Dubinski's (1994) prediction for the distribution of halo
shapes that will be produced in a cold dark matter cosmology.

\section*{acknowledgments}
 
We would like to thank Andy Newsam, Irini Sakelliou, Konrad Kuijken,
Marc Kamionkowski and Jacqueline van Gorkom for useful discussions. We
are also very grateful to the referee, James Binney, for his major
contribution to the clarity of this paper.

\end{document}